\documentclass[journal=jacsat,manuscript=article]{achemso}

\usepackage{amsmath,amsfonts,amsthm,amssymb,array,bm,dcolumn,enumerate,flafter,float,graphics,latexsym,makeidx,multirow,natbib,verbatim,xfrac}
\usepackage{graphicx}
\usepackage[english]{babel}
\usepackage[mathscr]{euscript}
\usepackage[utf8]{inputenc}
\usepackage{colortbl}
\usepackage{chemfig}
\usepackage[pdftex,colorlinks=true,linkcolor=blue,urlcolor=blue,citecolor=blue]{hyperref}

\usepackage[version=3]{mhchem} 

\author{Ana Luiza Costa Silva}
\email{analuiza@df.ufscar.br}
    \affiliation{Departamento de Física, Universidade Federal de São Carlos, 13565-905, São Carlos, SP, Brazil}

\author{Rafael Schio Wengenroth Silva}
\affiliation{Departamento de Física, Universidade Federal de São Carlos, 13565-905, São Carlos, SP, Brazil}

\author{Lucas Augusto Moisés}
\affiliation{Departamento de Física, Universidade Federal de São Carlos, 13565-905, São Carlos, SP, Brazil}

\author{Adenilson José Chiquito}
\affiliation{Departamento de Física, Universidade Federal de São Carlos, 13565-905, São Carlos, SP, Brazil}

\author{Marcio Peron Franco de Godoy}
\affiliation{Departamento de Física, Universidade Federal de São Carlos, 13565-905, São Carlos, SP, Brazil}

\author{Fabian Hartmann}
\affiliation{Julius-Maximilians-Universität Würzburg, Physikalisches Institut and Würzburg-Dresden Cluster of Excellence ct.qmat, Lehrstuhl für Technische Physik, Am Hubland, 97074 Würzburg, Deutschland}

\author{Victor Lopez-Richard}
\affiliation{Departamento de Física, Universidade Federal de São Carlos, 13565-905, São Carlos, SP, Brazil}

\title[An \textsf{achemso} demo]
  {From Memory Traces to Surface Chemistry: Decoding REDOX Reactions}

\abbreviations{IR,NMR,UV}
\keywords{American Chemical Society, \LaTeX}

\begin{document}

\begin{tocentry}

\includegraphics{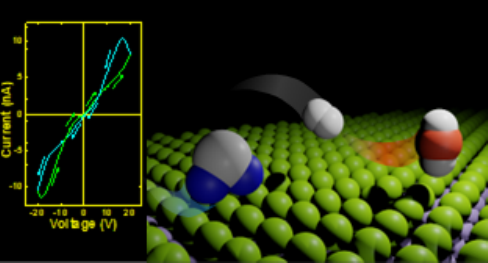}

\end{tocentry}

\begin{abstract}

Gas and moisture sensing devices leveraging the resistive switching effect in transition metal oxide memristors promise to revolutionize next-generation, nano-scaled, cost-effective, and environmentally sustainable sensor solutions. These sensors encode readouts in resistance state changes based on gas concentration, yet their nonlinear current-voltage characteristics offer richer dynamics, capturing detailed information about REDOX reactions and surface kinetics. Traditional vertical devices fail to fully exploit this complexity. This study demonstrates planar resistive switching devices, moving beyond the Butler-Volmer model. A systematic investigation of the electrochemical processes in Na-doped ZnO with lateral planar contacts reveals intricate patterns resulting from REDOX reactions on the device surface. When combined with advanced algorithms for pattern recognition, allow the analysis of complex switching patterns, including crossings, loop directions, and resistance values, providing unprecedented insights for next-generation complex sensors.

\end{abstract}

\section{Introduction}

The demand for advanced, miniaturized, and efficient information processing and sensing technologies is driving the development of innovative semiconductor materials beyond silicon~\cite{wang2021, wei2022, lee2021}. While some materials offer advantages in processing complexity, cost-effectiveness, flexibility, and environmentally sustainable, other materials enable applications that cannot be realized within silicon technology, e.g. spintronics, Mottronics, and in-memory computing~\cite{joksas2022, scheiderer2018, chiu2023} to name a few. One specific notable phenomenon in electronic devices, absent in silicon, is the resistive memory effect, which can be observed in materials fabricated with a metal-insulator-metal (MIM) structure, whose insulating layer is often composed of a transition metal oxide or a semiconductor oxide~\cite{datye2020, zhang2019, isyaku2021, lee2023}. Such structures can be utilized in traditional computing architectures and in novel beyond von Neumann computational architectures, such as in-memory computing, artificial neural networks or reservoir computing~\cite{zhang2024, abunahla2020, zhong2021}. Exploring the electronic conduction mechanisms of materials exhibiting memory traces also enable novel smart solutions for sensing applications of gases, gas mixtures and atmospheres.

Additional research indicates that certain semiconductor oxides exhibit resistive memory effects alongside with sensing functionalities, commonly referred to as ``gasistors''~\cite{lee2023, chae2023}. Among various approaches, elucidating the role of the surface in these processes presents a significant challenge. Studies on gas sensing underscore the critical importance of understanding the interactions between gas-sensitive materials and the ambient atmosphere~\cite{blackman2021}. However, the conventional MIM electrical contact configuration limits the detection capabilities of memristor-based gas sensors, as the gas primarily interacts with the upper electrode rather than directly with the oxide surface. Nontraditional setups such as lateral planar electrical contacts, provide deeper insights into the factors influencing the resistive memory effect. It also allows to expand the active detection area exposing the material surface either partially or completely to the target element~\cite{devi2024}. Thus, employing a lateral planar electrical contact geometry enables the investigation of the active role of the film surface for memory effects of semiconductor oxides when exposed to various atmospheric conditions that electrochemically react with it.  

We present both theoretical and experimental evidence of complex patterns emerging in the current-voltage loops of planar memristive devices. These phenomena result from REDOX reactions at various sites and under different environments, revealing a complexity that extends beyond the traditional Butler-Volmer model. Based on Na-doped ZnO planar memristive devices synthesized by spray-pyrolysis, we show experimentally how the current-voltage characteristics alter in the presence of different gas environments and how intricate behaviors beyond the resistance change emerge. These behaviors differ in the number of crossings, the symmetry of the response, and the polarity-dependent memory content. We derive an analytical solution by adapting the Butler-Volmer model, which correlates changes in the current-voltage response with the reaction type, the surrounding environment, and the specific location of the reaction site. 


\section{Methods}

\textbf{Sample Preparation.} The investigated system consists of  (002) oriented polycrystalline Na-doped ZnO thin films synthesized by the spray pyrolysis technique. This process is based on spraying a precursor solution, carried by dry air, onto a preheated substrate where the chemical-physical growth occurs. In this study, zinc acetate dihydrate (Zn(C$_{2}$H$_{3}$O$_{2}$)$_{2}$·2H$_{2}$O by Synth) and sodium hydroxide (NaOH by NEON) were used as the precursors. The zinc and sodium precursor solutions were prepared in distilled water with a molarity of $5\cdot 10^{-3} M$ and thermally stirred at 100 $^\circ$C. A nominal content of 10\%Na was used for Zn doping, referred to as ZnO:Na10\%. The deposition and growth of the films were carried out at 300 $^\circ$C with a flow rate of 0.3 mL/min. For further details on the sample preparation via spray pyrolysis, refer to \cite{Godoy20}.

\textbf{Material Characterization.} The Thermo Scientific K-Alpha spectrometer provided the X-ray photoelectron spectroscopy (XPS) for surface chemical analysis using the Al-K$\alpha$ monochromatic radiation (1486.6 eV). The spectra were fitted by a combination of Gaussian (70\%) and Lorentzian (30\%) functions employing the Shirley method as the baseline and aligning spectra on the adventitious carbon peak (C 1s – 284.8 eV) to correct charging effects. Film morphology was analyzed using a scanning electron microscope (SEM) model JEOL JSM 6510, equipped with a 20 kV electron beam and secondary electron (SE) mode. With a high conductive silver paint 503,  two planar electrical contacts separated by 1 mm were prepared, allowing a large area for atmosphere contact. For electrical characterization, a Keysight B2901A source measure unit provided the Current-voltage (I-V) curves and hysteresis loops with the sample mounted in a Linkam Scientific HFS600E cryostat equipped with two tungsten positional probes. All measurements were performed at room temperature under various atmospheric conditions, including ambient air, vacuum, controlled relative humidity (RH), and atmospheres of oxygen (O$_{2}$) and carbon dioxide (CO$_{2}$).

\textbf{Theoretical Methods.} For the theoretical simulation of the electrochemical reactions occurring at the surface under lateral bias, we propose an alternative to the conventional Butler-Volmer model. Our approach accounts for the complex interplay between different electrochemical processes at various surface sites, including their respective REDOX characteristics and environments. The electron transfer for each process is emulated by solving a set of differential equations grounded in the relaxation time approximation. These equations describe the time evolution of surface electron density fluctuations due to local reactions. Numerical solutions were obtained using standard Runge-Kutta methods, which allowed us to model the dynamic conductance response under varying voltage sweeps and simulate the resulting current-voltage hysteresis loops. This approach provides a nuanced view of concomitant electrochemical processes in laterally biased device configurations.

\section{Results and discussion}

\begin{figure}
    \centering
   \includegraphics[width=1\textwidth]{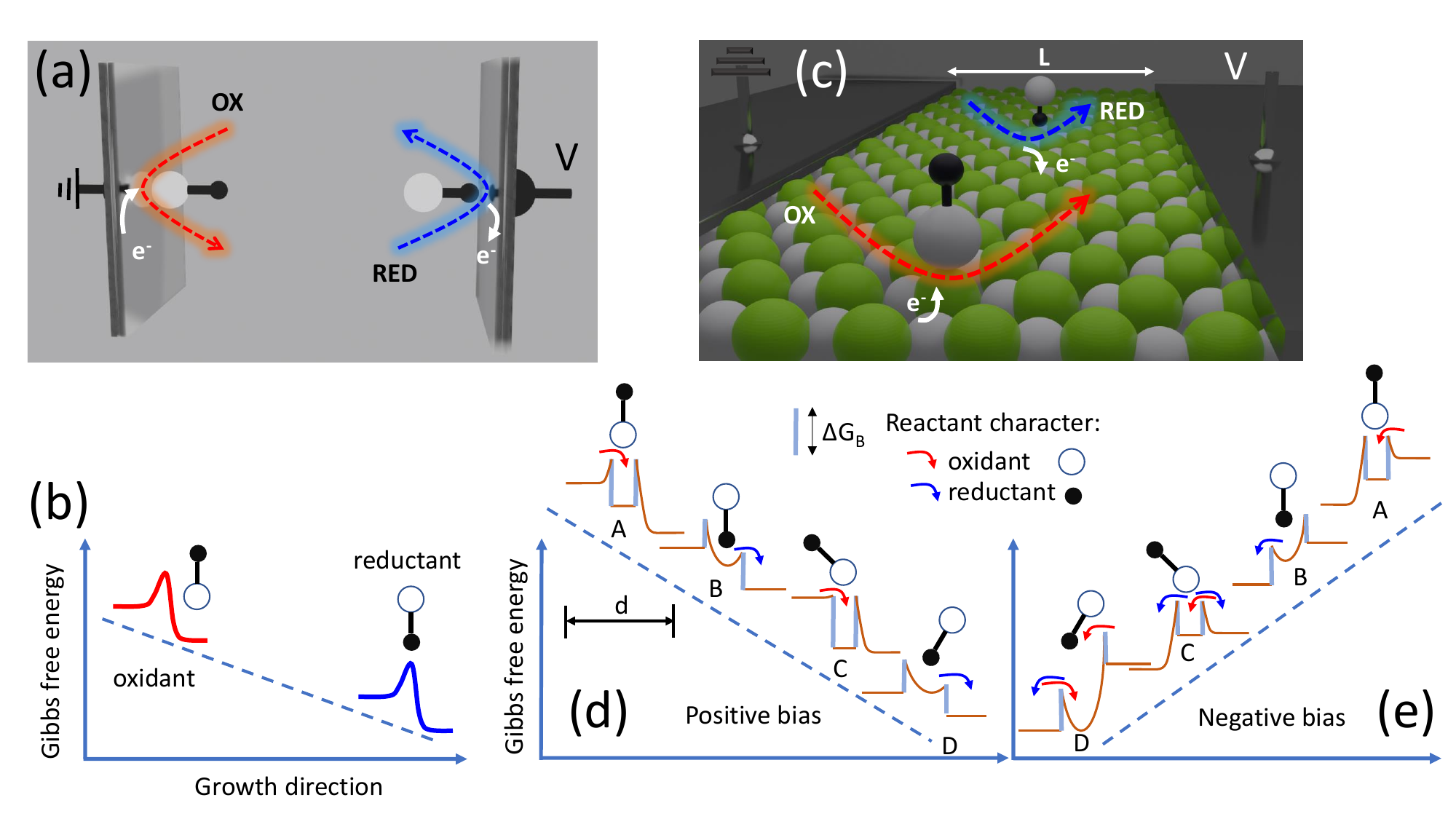}
    \caption{(a) Schematic representation of REDOX reactions at perpendicularly biased electrodes and (b) the corresponding Gibbs energy profile close to the surfaces. (c) Representation of the REDOX reactions under lateral biasing, (d) and (e) potential Gibbs energy profiles for lateral electron activation and trapping under positive and negative biases, respectively: A, predominant surface oxidation for a symmetric molecular adsorption; B, predominant surface reduction for a symmetric molecular adsorption; C, predominant surface oxidation (reduction) under positive (negative) bias; D, predominant surface reduction (oxidation) under positive (negative) bias.}
    \label{fig1}
\end{figure}

The direction and rate of electrochemical reactions depend on the combined influence of thermodynamics and kinetics. Thermodynamics, captured by the Gibbs free energy change ($\Delta G_B$), dictates the spontaneity and equilibrium of the reaction~\cite{landau1980}. A negative $\Delta G_B$ indicates a thermodynamically favorable reaction, while the magnitude of $\Delta G_B$ reflects the driving force. However, $\Delta G_B$ alone doesn't provide information about the reaction time and a kinetic approach is thus needed. In general, the activation energy over an energy barrier governs the reaction rates. In electrochemical reactions, electron transfer processes involve overcoming energy barriers at the electrode surface, as illustrated in Figure~\ref{fig1}. Figure~\ref{fig1}(a) depicts the schematic representation of REDOX reactions at perpendicularly polarized electrodes. The Butler-Volmer equation serves as a bridge between thermodynamic and kinetic reactions, specifically those occurring at the electrode-electrolyte interface~\cite{bard2001}. It relates the carrier flux or electron transfer rate, $f_{BW}$, to the difference between the actual electrode applied potential and the equilibrium potential predicted by thermodynamics, $\Delta \phi$, and can be expressed as~\cite{atkins2006}
\begin{equation}
f_{BW} = f_0 \left\{ \exp \left[ \alpha  \frac{e \Delta \phi}{k_BT} \right] - \exp \left[ -\left(1-\alpha\right) \frac{e \Delta \phi}{k_BT} \right] \right\}.
\label{BW}
\end{equation}
Here, $\Delta \phi$ is the potential difference at the electrode-insulator interface and $f_0$ relates to exchange current density which is proportional to the equilibrium activation rate, $k e^{-\Delta G_B/k_B T}$.
$k$ is a proportional constant, and $k_BT$ the thermal energy. Figure~\ref{fig1}(b) schematically depicts the Gibbs energy profile near the surfaces for perpendicularly polarized electrodes. The probability of an electron overcoming the activation barrier determines the reaction rate and is influenced by the asymmetry of the barriers for the forward and reverse reactions. It is characterized by the transfer coefficient $\alpha \in [0,1]$ in Equation~\ref{BW}. A transfer coefficient of $\alpha=0.5$ represents a symmetrical reaction mechanism, where both forward and reverse reactions have the same activation barriers and proceed at equal rates.

The Butler-Volmer equation has limitations in accurately describing electrochemical reactions at laterally biased surfaces due to several factors. It does not explicitly consider the potential concomitance of oxidation and reduction reactions occurring simultaneously at surface defects, such as vacancies or dangling bonds. The equation also assumes a uniform and homogeneous electrode surface, whereas laterally biased surfaces may exhibit gradients in properties such as work function or electron density. For instance, surface defects and dangling bonds can act as active sites, displaying lower activation energies for oxidation and reduction reactions compared to defect-free regions. Consequently, at these defect sites, both oxidation and reduction reactions associated with a specific reactant compete, influencing the overall electrical conductivity.

Let's consider the case of contacts placed laterally on an electrode as shown in Figure~\ref{fig1}(c), a configuration relevant to many technologies like finger-contact sensors and our device. These contacts experience a potential difference, leading to a linear gradient of the electrochemical potential across the electrode surface, as sketched in Figure~\ref{fig1}(d) and (e) for positive and negative bias, respectively. Let's also assume, as represented in that panel, that the surface contains active sites of size $d$, resulting in a local potential drop of $|\eta| V$, where $|\eta| = d/L$. We can then consider various configurations of potential profiles to characterize the electrochemical process triggered at these sites, describing how different chemical species, with either oxidant or reductant character, interact with the surface. The electron transfer rate for all these configurations can be expressed by the following equation as described in Ref.~\citenum{LopezRichard2022}

\begin{equation}
f_{L}(V) = \frac{f_0}{\eta} \left\{ \exp \left[ \alpha \eta \frac{e V}{k_BT} \right] + \exp \left[ -\left(1-\alpha \right) \eta \frac{eV}{k_BT} \right] - 2\right\}.
\label{lat}
\end{equation}
In this equation, $f_0 \propto \exp(-\Delta G_B/k_B T)$, where $\Delta G_B$ is the Gibbs activation energy barrier represented by the gray lines in Figure~\ref{fig1}(d) and (e). More information on the details of this equation and deviations from the Butler-Volmer model can be found in the Suppl. material. The transfer coefficient $\alpha$ in Eq.~\ref{lat} continues to weight the configuration symmetry, and $\eta$ corresponds to the reaction character: for $\eta<0$ the reaction is reductant and otherwise oxidant.

In Figure~\ref{fig1}, the red and blue arrows represent the main electron fluxes at the interfaces of the active sites, corresponding to electron capture and release due to the REDOX character, respectively. In symmetric cases ($\alpha=0.5$), two primary configurations emerge: one characterized by a purely oxidant contribution independent of polarity, denoted as case A and another (case B) featuring a purely reductant character, also independent of polarity. Additionally, two asymmetric configurations, labeled as C and D, are depicted to underscore the polarity dependence of the REDOX reactions in those cases. 

\begin{figure}[!htb]
\centering
  \includegraphics[width=1\textwidth]{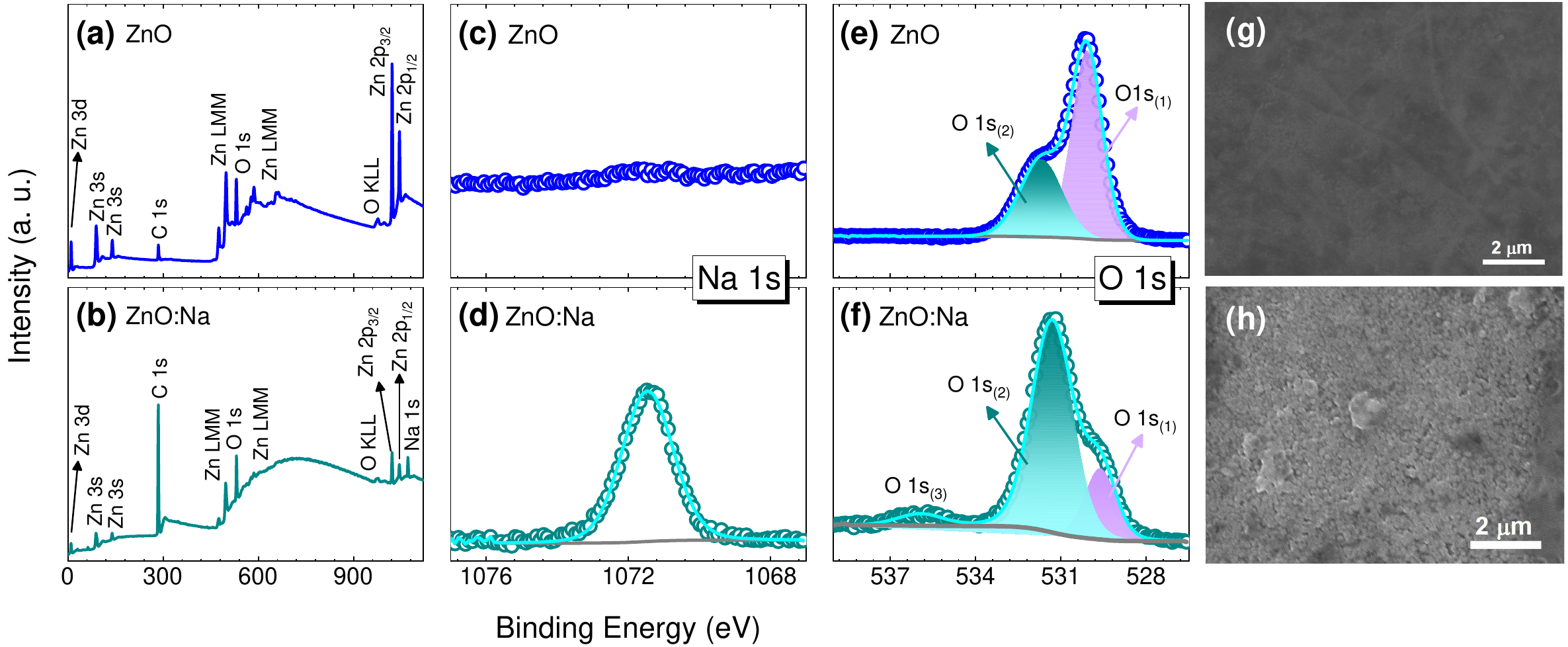}
     \caption{XPS spectra of ZnO and ZnO:Na films -- (a) and (b) survey scans, (c) and (d) high-resolution XPS of the Na 1s level, (e) and (f) high-resolution XPS of the O 1s level. The SEM images at 2 $\mu$m scale with 750x magnification are shown in (g) and (h) for the ZnO and ZnO:Na, respectively.}
    \label{fig2}
\end{figure}

Surface REDOX reactions can be significantly enhanced by introducing reactive elements into oxide hosts. For example, the spray-pyrolysis technique offers an efficient and versatile method for doping ZnO with strategic elements. This approach is cost-effective, operates without the need for vacuum conditions, and supports large-scale material production. Furthermore, the residual gases released during film growth are environmentally benign, making this technique both practical and eco-friendly. The experiments were conducted with resistive memory devices based on Na-doped ZnO, prepared as described in Ref.~\citenum{silva2024} with 10\% nominal Na-content. Insights about the surface feature, like surface defects and adsorption sites, were obtained by X-ray photoelectron spectroscopy (XPS) as shown in Figure~\ref{fig2}. Figures~\ref{fig2}(a) and (b) display the survey scans for the undoped ZnO (reference sample) and Na-doped ZnO, respectively. The data show that the surfaces primarily consist of Zn, O, and C, with Na detected only in the Na-doped sample -- indicating the absence of contamination in the growth process as depicted in the undoped ZnO film, Figure~\ref{fig2}(c), while the ZnO:Na sample exhibits a binding energy at 1071.4 eV -- attributed to the Na–O bond~\cite{erdogan21}, Figure~\ref{fig2}(d).

The high-resolution spectra in the O 1s are shown in Figures~\ref{fig2}(e) e (f) for ZnO and ZnO:Na, respectively. Two distinct peaks are observed: the lower energy peak, O 1s$_{(1)}$, corresponds to oxygen bonded to Zn or substitutional Na in the ZnO wurtzite structure~\cite{Ye17,mueen20}, while the higher energy peak,  O 1s$_{(2)}$, can be attributed to two possible surface species -- oxygen vacancies (V$_{O}$) in the ZnO lattice~\cite{erdogan21,guo15} or hydroxyl groups (OH) from chemisorbed water~\cite{frankcombe2023,idriss2021}. Furthermore, the Na-doped ZnO sample exhibits a higher energy shoulder at 535.9 eV (O 1s$_{(3)}$) assigned to water molecules in the gas phase interacting with the atmosphere \cite{yamamoto08}. Notably, the higher intensity of O 1s$_{(2)}$ indicates a Na-inducing mechanism that increases the density of oxygen vacancies\cite{silva2024giant}, which are then partially occupied by OH groups detected by XPS. Figure~\ref{fig2} also presents SEM images illustrating the surface morphology of both undoped and Na-doped ZnO films at a 2~$\mu$m scale. In Figure~\ref{fig2}(g), the undoped ZnO shows a smooth, homogeneous surface, typical of nanostructured polycrystalline films. The addition of sodium, however, introduces distinct morphological features, such as micro- and nanostructures. Figure~\ref{fig2}(h) reveals the nanoporous surface morphology of Na-doped ZnO, whose hydrophilic nature can trap water molecules, evidenced by the O 1s$_{(3)}$ band in the XPS spectrum. These microscopic observations underscore the significant impact of the synthesis method on defining structural characteristics, particularly in relation to dopant effects on surface morphology and potential applications. Dopant incorporation influences electrochemical properties by modifying the active surface area available for reactions, which may lead to non-uniform charge distribution and fluctuations in electric fields and reaction rates \cite{wu2022, joshi2020}. Such effects can result in unregulated or slower REDOX processes and the formation of charge traps that impede electron transfer. Furthermore, dopant agglomerations impair ionic mobility, which compromises the efficiency and reliability of devices such as sensors and resistive memory systems, ultimately reducing their sensitivity, stability, and overall performance \cite{rehman2019,li1999, wang2010}.

\begin{figure}[!htb]
\centering
 \includegraphics[width=0.5\textwidth]{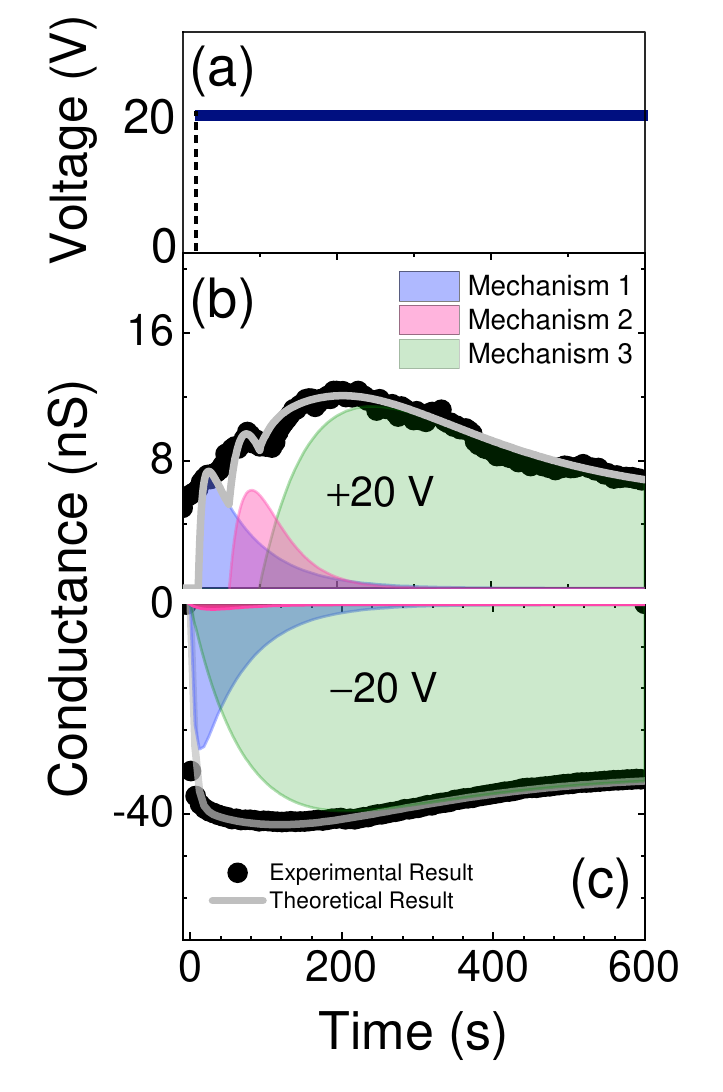}
    \caption{(a) Rectangular voltage pulses for which the conductance evolution with time was obtained: (b) for positive bias and (c) for negative bias.}
\label{fig3}
\end{figure}

Experiments using normal pulse voltammetry with rectangular voltage pulses, as depicted in Figure~\ref{fig3}(a), were conducted in an ambient atmosphere. Figures~\ref{fig3}(b) and (c) depict the observed evolution of the conductance during the application of positive and negative bias voltage pulses, respectively, with a pulse duration of 600 seconds. Our analysis reveals at least six distinct contributions with contrasting timescales. To facilitate the fitting process, we combine these contributions into pairs, designated as mechanisms 1, 2, and 3, as illustrated in Figures~\ref{fig3}(b) and (c). For mechanism 1, two characteristic timescales can be extracted: $\tau_{1}$ = 5 s and $\tau_{2}$ = 63 s. Mechanism 2 exhibits timescales of $\tau_{1}=25.8 s$ s and $\tau_{2}=33.2 s$. Finally, the mechanism 3 displays the longest duration with $\tau_{1}$ = 90 s and $\tau_{2}$ = 155 s. Nonmonotonic temporal transients in the conductance can indicate the concurrence of processes occurring at different timescales.

\begin{figure}[!htb]
    \centering
   \includegraphics[width=1.0\textwidth]{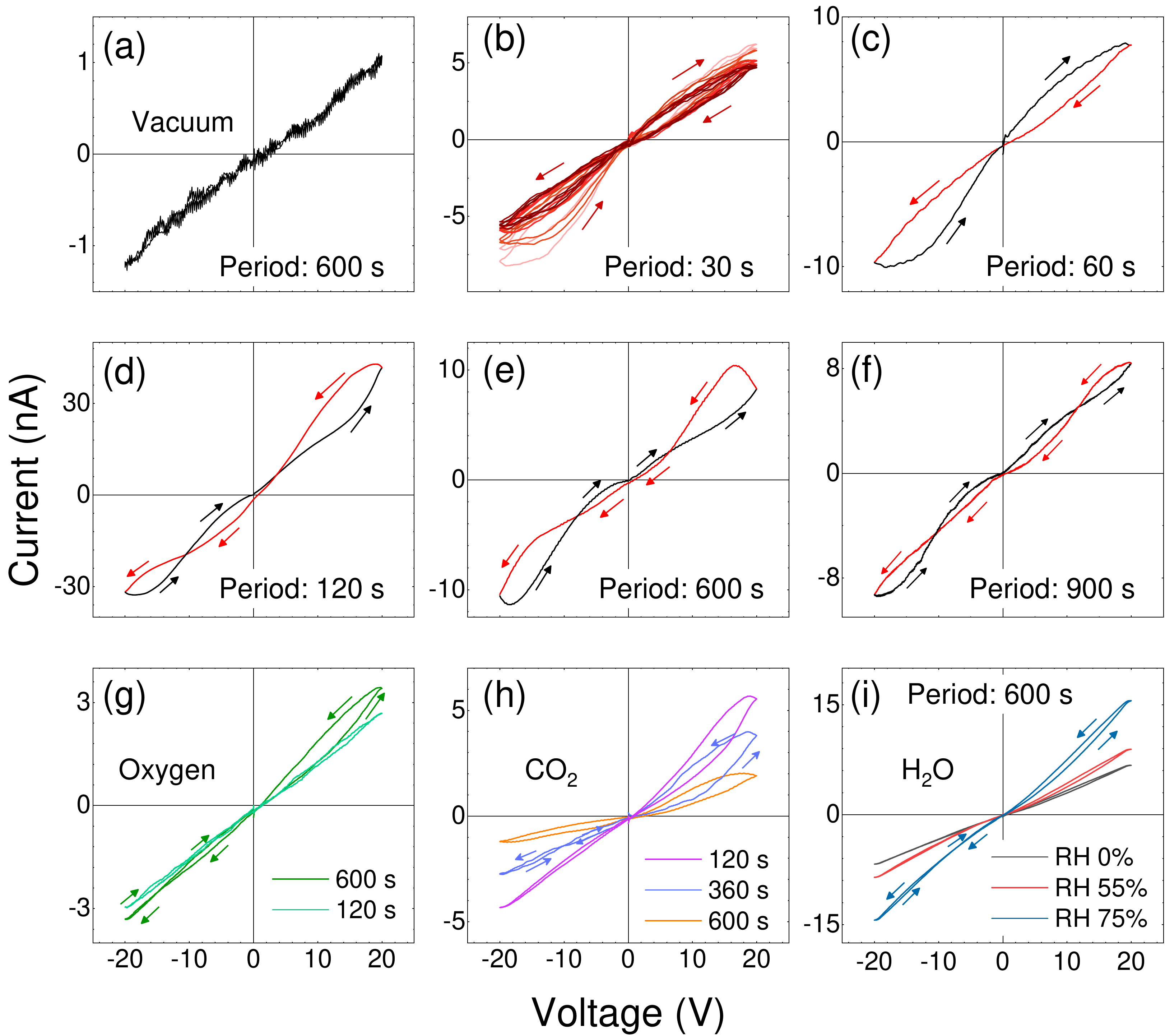}
    \caption{(a) Experimental current-voltage sweep performed in vacuum, (b) current-voltage loops performed over a 30-second period, illustrating the stability trend of the system through the cycles. (c)-(f) Stable current-voltage loops performed in ambient atmosphere for increasing voltage sweeping periods. Stable loops obtained under controlled atmospheres of (g) oxygen (O$_{2}$) and (h) carbon dioxide (CO$_{2}$). Relative humidity of: 0\% (dry air), 55\% $\pm$ 5\% and 75\% $\pm$ 5\% were also analyzed at scan periods of (i) 600 seconds.}
    \label{fig4}
\end{figure}

The experimental cyclic voltammetry analysis was conducted using voltage sweeps with a 20 V amplitude. The results, displayed in Figure~\ref{fig4}, show current-voltage loops under different scanning periods. In vacuum, as shown in Figure~\ref{fig4}(a), the measurement exhibits ohmic behavior, with no observable hysteresis. In contrast, under ambient atmosphere (Figures~\ref{fig4}(b)-(f)) memory traces appear which are sensitive to the sweep periods. First, Figure~\ref{fig4}(b) shows the evolution of the hysteresis loops towards stabilization for a scanning period of 30 seconds. All other figures correspond to stabilized responses. A complex pattern with different maximal and minimal currents, memory content, and number of crossings emerges under different periods of the voltage driving. While for 60s only a single crossing at zero bias emerges, the number of crossing increases to 2 for larger periods with an avoided crossing at zero bias. Also, the maximal and minimal current values exhibit a non-monotonic behavior with maximal currents observed at a period of 120s.

To examine how complex patterns emerge depending on the REDOX character, Figures~\ref{fig4}(g)-(i) present the current-voltage hysteresis loops generated under controlled atmospheres of O$_2$, CO$_2$, and H$_2$O (measured at varying relative humidity (RH) levels). The conductance in all three environments is significantly higher than in vacuum, with distinct differences observed between the atmospheres. One notable difference is the time scale, with CO$_2$ reactions responding more rapidly than those in O$_2$ or H$_2$O. It is clear that the complex effects observed under ambient conditions cannot be attributed solely to the sum of these three components. The intricate dynamics revealed in Figures~\ref{fig4}(b)-(f) likely result from the combined electrochemical effects of various species, potentially including interactions with CO molecules present in the atmosphere~\cite{blackman2021}, which were not examined in this study. Additionally, moisture is expected to have complex effects~\cite{Milano2021}, potentially involving ionic transport channels through surface vacancies and protonic diffusion over hydroxyl groups~\cite{Messerschmitt2015}. The formation of adsorbed carbonates on the ZnO surface hydroxyls cannot be ruled out either~\cite{gankanda16,kahyarian17}.

Pattern recognition is an area where artificial intelligence significantly surpasses traditional computing architectures, including quantum computing~\cite{sheridan2014}. As shown in Figure \ref{fig4}, the hysteresis loops differ in several key aspects: the number of crossings, the minimum and maximum current values, the memory content within the loops, and the symmetry of the current-voltage response under voltage polarity reversal. Traditionally, assessments of device sensitivity have focused on variations between the Low Resistance State (LRS) and High Resistance State (HRS) in response to different gases and concentrations 
While this approach facilitates straightforward integration with readout circuits, it overlooks the much richer and more selective information encoded in the full loop dynamics. For example, under ambient conditions, multiple crossings are evident, whereas a pure O$_2$ atmosphere shows no such crossings, and a CO$_2$ atmosphere exhibits multiple crossings during specific scanning intervals. Additionally, the symmetry of the response allows for clear differentiation between O$_2$ and CO$_2$ atmospheres. In contrast, high humidity levels result in a weaker memristive response but larger minimum and maximum currents. This analysis reveals distinct and identifiable patterns corresponding to different atmospheric conditions, enabling precise differentiation between them.

The origin of the complex behavior can be attributed to the simultaneous occurrence of different electrochemical reactions at surface sites with varying environments. These reactions induce fluctuations in surface electron density, $\delta n$, leading to deviations of the surface conductance from its equilibrium value, $G_0$, by $G_n \propto \delta n$. To fully account for the contributions of each electrochemical reaction, the electron transfer rate for each reaction type $(j)$ must be considered. This requires introducing the index $(j)$ in Eq.~\ref{lat}, which identifies each process and its corresponding transfer rate, $f_{L}^{(j)}$, reflecting different environments ($\Delta G_B^{(j)}$ and $\alpha^{(j)}$) and REDOX characteristics ($\eta^{(j)}$). Each process, governed by its relaxation time ($\tau_{j}$), evolves over according to
\begin{equation}
    \frac{d \delta n^{(j)}}{dt}=-\frac{\delta n^{(j)}}{\tau_{j}}+ f_{L}^{(j)}(V),
    \label{rate}
\end{equation}
resulting in a total fluctuation $\delta n (V)=\sum_j \delta n^{(j)}$ which influences the conductance as 
\begin{equation}
    G(V)= G_0+\frac{e \mu}{L^2} \delta n (V),
    \label{cond}
\end{equation}
where the current is defined by $I=G\cdot V$, with $\mu$ representing electron mobility and $e$ the elementary charge. Solving Equation~\ref{rate} under this condition yields the electron fluctuation, $\delta n^{(j)}(t)=\tau_{j}f_{L}^{(j)}(V_0)+\left[\delta n^{(j)}(0)-\tau_{j}f_{L}^{(j)}(V_0) \right]\exp(-t/\tau_{j})$ and the corresponding conductance response to the voltage step (assumed to start at $t=0$)
\begin{equation}
    G=\tilde{G}_0+\sum_j \tilde{G}_{(j)} \exp\left(-\frac{t}{\tau_{j}}\right),
    \label{st}
\end{equation}
where $\tilde{G}_0=G_0+\frac{e\mu}{L^2}\sum_j \tau_{j}f_{L}^{(j)}(V_0)$ represents the background conductance contribution and $\tilde{G}_{(j)}=\frac{e \mu}{L^2}\left[\delta n^{(j)}(0)-\tau_{j}f_{L}^{(j)}(V_0) \right]$. Note that the sign of $\tilde{G}_{(j)}$ depends on the initial fluctuation condition and can be positive or negative. Beyond this ambiguity, Equation~\ref{st} captures the interplay between the influence of various types of active sites and their characteristic time constants, $\tau_{j}$. The experiments using normal pulse voltammetry with rectangular voltage pulses (see Figures~\ref{fig3}(b) and (c)) have been fitted using Equation~\ref{st}. While the model captures the core reaction kinetics, it doesn't account for mass transport and interdiffusion, which can introduce delays in the transient response of conductance. To address this and fit the experimental results in Figures~\ref{fig3}(b) and (c), a time delay ($t \rightarrow t - \Delta t^{delay}_{(j)}$) was incorporated into Equation~\ref{st}.

To replicate the intricate patterns observed in the experimental cyclic voltammetry data shown in Figure~\ref{fig4}, current-voltage curves can be calculated using Equations~\ref{rate} and~\ref{cond}. This involves applying various transfer rates, $f_{L}^{(j)}$, which account for different environments ($\Delta G_B^{(j)}$ and $\alpha^{(j)}$) and REDOX character ($\eta^{(j)}$). Our primary focus is on emulating the number of crossings, the symmetry of the memory content, maximal current values, and the loop direction, all of which are influenced by these variations. 

\begin{figure}[!htp]
    \centering
   \includegraphics[width=1\textwidth]{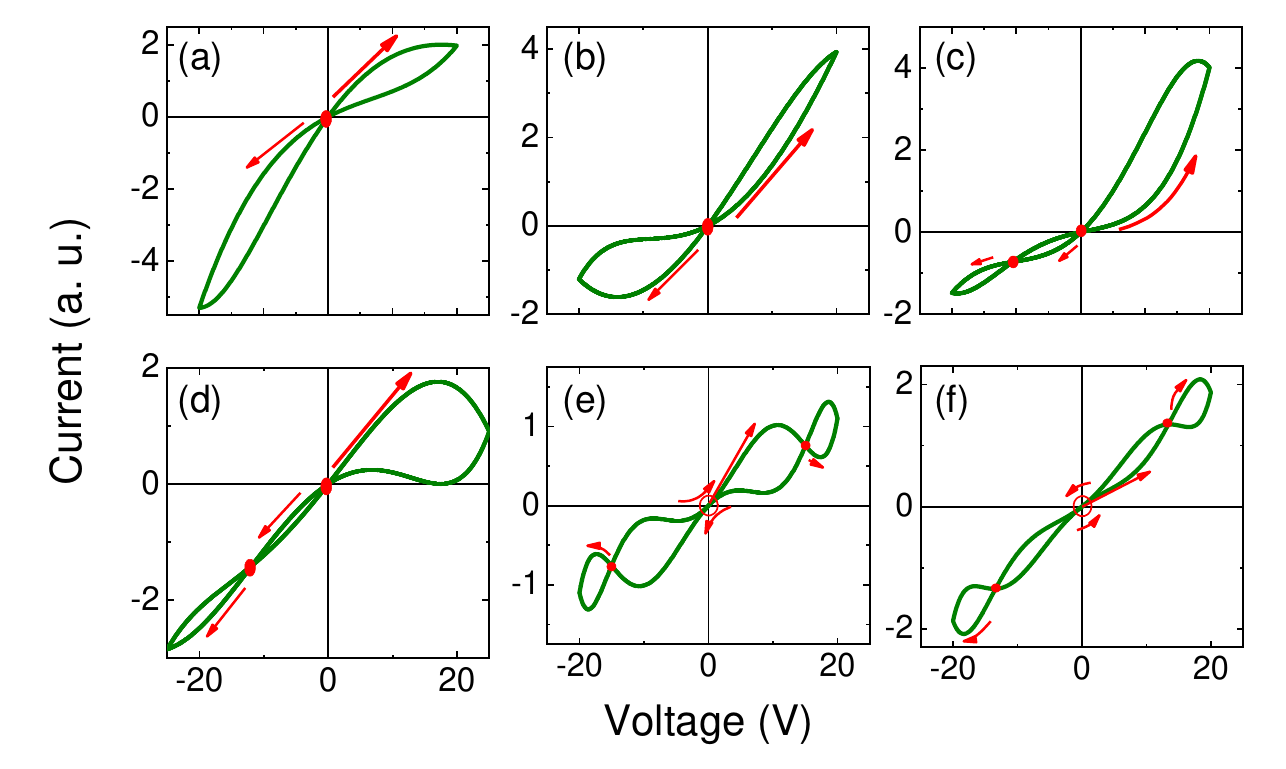}
    \caption{Theoretical  current-voltage loops: (a), (b), (c), and (d) for a single asymmetric electron transfer process, while (e) and (f) illustrate current-voltage loops for a combination of two symmetric electron transfer processes.}
    \label{fig5}
\end{figure}

First, we consider a simple single process that breaks inversion symmetry ($\alpha \rightarrow 1$), as illustrated in diagrams C or D of Figure~\ref{fig1}(d), and further depicted in Figure~\ref{fig5}(a) and (b). The resulting memory loops are asymmetric with a zero-voltage crossing. For an oxidizing reaction (Figure~\ref{fig5}(a)), the maximum current observed for positive polarity is lower than that for negative polarity, and the current-voltage loop exhibits a clockwise rotation. Conversely, for a reducing reaction (Figure~\ref{fig5}(b)), the maximum current for positive polarity exceeds that for negative polarity, and the loop rotates counterclockwise. 

For a more symmetrical case (while keeping $\alpha \neq 0.5$), the loop shape, even with a single transfer mechanism, may exhibit multiple crossings within the first or third quadrant of the current-voltage plane. Two possible scenarios are depicted in Figures~\ref{fig5}(c) and (d). It is important to note that the presence of an additional crossing in either the upper or lower loop is influenced by the transition in reaction character (from oxidant to reductant and vice versa) at positive or negative polarity, as well as by the voltage sweep rate. Further details can be found in the Supplementary Material and in Ref.~\citenum{LopezRichard2022}.

When two concurrent transfer mechanisms with similar relaxation times, such as those characterized by $\tau_2$ and $\tau_3$, are present, the resulting memory traces become even more intricate. In this situation, varying the voltage period can cause the total area of the upper loop to change sign, leading to a complete reversal of the loop direction and producing multiple crossings in the first quadrant~\cite{LopezRichard2022}.

A specific case involving two concurrent symmetric oxidation and reduction processes, each with identical relaxation times but differing activation energy barriers as described in Equation~\ref{lat}, is illustrated in Figure~\ref{fig5}(e). This scenario produces a pinched, non-crossing hysteresis at $V = 0$, due to the combined influence of the two opposing transfer functions. The transition between oxidant and reductant behavior with varying bias introduces additional crossings in the first and third quadrants. By adjusting the relative strengths of these two processes, the loop direction can be reversed, as depicted in Figure~\ref{fig5}(f). Breaking the inversion symmetry leads to polarity dependence, which results in a crossing at $V = 0$ and the possibility of an extra crossing in one of the quadrants. 

Our analysis reveals that at shorter periods (Figures \ref{fig4}(b) and (c)), the dynamics correspond to asymmetric adsorption, characterized by oxidation at positive polarity and reduction at negative polarity. This behavior is accurately captured by the theoretical model shown in Figure \ref{fig5}(a). As the period lengthens, an inverted dynamic emerges, characterized by multiple crossings, as depicted in Figure\ref{fig4}(d) and aligned with the theoretical patterns in Figure\ref{fig5}(c) or (e). At the longest periods, these asymmetric dynamics appear to subside, revealing the influence of two concurrent symmetric contributions—oxidation and reduction—which can be observed in Figure~\ref{fig4}(e). Here, surface reduction dominates at higher voltages, leading to curve intersections in both quadrants without crossing at the origin. The experimental curves in Figures~\ref{fig4}(e) and (f) show strong similarities with the theoretical predictions in Figure~\ref{fig5}(e), while those in controlled atmospheres align with the processes illustrated in Figures~\ref{fig4}(b) and (c). Table \ref{Tab1} summarizes the parameters extracted from these experiments, with the final three columns corresponding to the results of the computational simulations. Additional I-V loops for different sweep periods are presented in Figure S4 in the Supplementary Information.

\begin{table}[]
\scalebox{0.74}{
\centering
\caption{\label{Tab1} Parameters extracted from the I-V hysteresis curves. CW and CCW, denoting the direction of the loop under positive polarity—clockwise and counterclockwise, respectively. Y (Yes) and N (No) indicate the presence of zero-bias crossings, multiple crossings, symmetry of the current in the positive and negative quadrants, and symmetry of the hysteresis loop contents.The parameter $\eta$ characterizes the oxidizing or reducing nature of the process, while the mechanism denotes whether the process is a single asymmetric (1 AS) process or a combination of two symmetric processes (2 S). The final column shows which theoretical curve corresponds to the experimental data.}
\begin{tabular}{ccccccccc}
\hline
Period               & \begin{tabular}[c]{@{}c@{}}Loop\\ direction\end{tabular} & \begin{tabular}[c]{@{}c@{}}Zero\\ crossing\end{tabular} & \begin{tabular}[c]{@{}c@{}}Additional\\ crossing\end{tabular} & \begin{tabular}[c]{@{}c@{}}Symmetry\\ (current)\end{tabular} & \begin{tabular}[c]{@{}c@{}}Symmetry\\ (content)\end{tabular} & $\eta$                                                      & Mechanism            & \multicolumn{1}{l}{Figure 3} \\ \hline
                     & \multicolumn{5}{c}{\textbf{ambient  atmosphere}}                                                                                                                                                                                                                                                                 &                                                             &                      & \multicolumn{1}{l}{}         \\
30 s                 & CW                                                       & Y                                                       & N                                                             & Y                                                            & Y                                                            & positive                                                    & 1 AS                 & a                            \\
60 s                 & CW                                                       & Y                                                       & N                                                             & Y                                                            & Y                                                            & positive                                                    & 1 AS                 & a                            \\
120 s                & CCW                                                      & Y                                                       & Y                                                             & Y                                                            & N                                                            & positive                                                    & 1 AS                 & c                            \\
600 s                & CW/CCW                                                   & N                                                       & Y                                                             & Y                                                            & Y                                                            & \begin{tabular}[c]{@{}c@{}}negative\\ positive\end{tabular} & 2 S                  & e                            \\
900 s                & CW/CCW                                                   & N                                                       & Y                                                             & Y                                                            & Y                                                            & \begin{tabular}[c]{@{}c@{}}negative\\ positive\end{tabular} & 2 S                  & e                            \\
\multicolumn{1}{l}{} & \multicolumn{1}{l}{}                                     & \multicolumn{1}{l}{}                                    & \multicolumn{1}{l}{}                                          & \multicolumn{1}{l}{}                                         & \multicolumn{1}{l}{}                                         & \multicolumn{1}{l}{}                                        & \multicolumn{1}{l}{} & \multicolumn{1}{l}{}         \\
                     & \multicolumn{5}{c}{\textbf{in O$_{2}$ atmosphere}}                                                                                                                                                                                                                                                               &                                                             &                      &                              \\
\multicolumn{1}{l}{} & \multicolumn{1}{l}{}                                     & \multicolumn{1}{l}{}                                    & \multicolumn{1}{l}{}                                          & \multicolumn{1}{l}{}                                         & \multicolumn{1}{l}{}                                         & \multicolumn{1}{l}{}                                        & \multicolumn{1}{l}{} & \multicolumn{1}{l}{}         \\
120 s                & CCW                                                      & Y                                                       & N                                                             & Y                                                            & N                                                            & negative                                                    & 1 AS                 & b                            \\
600 s                & CCW                                                      & Y                                                       & N                                                             & Y                                                            & N                                                            & negative                                                    & 1 AS                 & b                            \\
\multicolumn{1}{l}{} & \multicolumn{1}{l}{}                                     & \multicolumn{1}{l}{}                                    & \multicolumn{1}{l}{}                                          & \multicolumn{1}{l}{}                                         & \multicolumn{1}{l}{}                                         & \multicolumn{1}{l}{}                                        & \multicolumn{1}{l}{} & \multicolumn{1}{l}{}         \\
                     & \multicolumn{5}{c}{\textbf{CO$_{2}$  atmosphere}}                                                                                                                                                                                                                                                              &                                                             &                      &                              \\
\multicolumn{1}{l}{} & \multicolumn{1}{l}{}                                     & \multicolumn{1}{l}{}                                    & \multicolumn{1}{l}{}                                          & \multicolumn{1}{l}{}                                         & \multicolumn{1}{l}{}                                         & \multicolumn{1}{l}{}                                        & \multicolumn{1}{l}{} & \multicolumn{1}{l}{}         \\
120 s                & CCW                                                      & Y                                                       & N                                                             & Y                                                            & N                                                            & negative                                                    & 1 AS                 & b                            \\
360 s                & CCW                                                      & Y                                                       & Y                                                             & N                                                            & N                                                            & positive                                                    & 1 AS                 & c                            \\
600 s                & CCW                                                      & Y                                                       & N                                                             & N                                                            & N                                                            & negative                                                    & 1 AS                 & b                            \\
\multicolumn{1}{l}{} & \multicolumn{1}{l}{}                                     & \multicolumn{1}{l}{}                                    & \multicolumn{1}{l}{}                                          & \multicolumn{1}{l}{}                                         & \multicolumn{1}{l}{}                                         & \multicolumn{1}{l}{}                                        & \multicolumn{1}{l}{} & \multicolumn{1}{l}{}         \\
                     & \textbf{}                                                &                                                         & \textbf{H$_{2}$O atmosphere}                                   &                                                              &                                                              &                                                             &                      &                              \\
\multicolumn{1}{l}{} & \multicolumn{1}{l}{}                                     & \multicolumn{1}{l}{}                                    & \multicolumn{1}{l}{}                                          & \multicolumn{1}{l}{}                                         & \multicolumn{1}{l}{}                                         & \multicolumn{1}{l}{}                                        & \multicolumn{1}{l}{} & \multicolumn{1}{l}{}         \\
600 s - RH 0\%       & CCW                                                      & Y                                                       & N                                                             & Y                                                            & N                                                            & negative                                                    & 1 AS                 & b                            \\
600 s - RH 55\%      & CCW                                                      & Y                                                       & N                                                             & Y                                                            & N                                                            & negative                                                    & 1 AS                 & b                            \\
600 s - RH 75\%      & CCW                                                      & Y                                                       & Y                                                             & Y                                                            & N                                                            & positive                                                    & 1 AS                 & c                            \\ \hline
                     &                                                          &                                                         &                                                               &                                                              &                                                              &                                                             &                      &                             
\end{tabular}}
\end{table}

\section{Conclusions}

In conclusion, our study uncovers the intricate dynamics of resistive memory in Na-doped ZnO thin films under diverse atmospheric conditions. The complex current-voltage hysteresis loops observed emphasize the influence of asymmetric adsorption configurations, the significance of voltage sweep rates, and the local environment in shaping memory traces. These rich dynamics arise from the combined electrochemical interactions of multiple species. Our proposed transfer rate formula, specifically designed for lateral biasing, offers a powerful tool for analyzing electrochemical reactions at surface states. By integrating this refined model, we can achieve a more precise characterization of the dynamic processes governing resistive memory effects, leading to a deeper understanding of surface interactions and transfer mechanisms. This approach opens new avenues for optimizing electrochemical systems across various applications.

\newpage

\begin{acknowledgement}
This study was financed in part by the Coordenação de Aperfeiçoamento de Pessoal de Nível Superior - Brazil (CAPES); the Conselho Nacional de Desenvolvimento Científico e Tecnológico - Brazil (CNPq) Proj. 311536/2022-0 and 312254/2023-7; and FAPESP: Procs. 2023/17490-2 and 2023/05436-3. This research used resources provided by the National Nanotechnology Laboratory (LNNano), which operates within the Brazilian Center for Research in Energy and Materials (CNPEM), a private non-profit organization supervised by the Ministry of Science, Technology, and Innovations (MCTI) of Brazil. We are grateful to the XPS team for their help during the experimental work (proposals XPS-28104 and XPS-28108). 

\end{acknowledgement}

\bibliography{PaperMemory.bbl}

\end{document}